\documentclass{aa}

\usepackage[varg]{txfonts}
\usepackage{graphicx}
\usepackage{natbib}
\bibpunct{(}{)}{;}{a}{}{,} 
\usepackage{amssymb}
\usepackage{amsmath}
\usepackage{multirow}
\usepackage{appendix}
\usepackage[colorlinks=true,     linkcolor=blue, citecolor=blue, filecolor=blue, urlcolor=blue]{hyperref}

\begin{document}

\title{Detection of wave activity within a realistic 3D MHD quiet sun simulation}

\author{George Cherry\thanks{\email{georgche@uio.no}}\inst{1,2}  \and Boris Gudiksen\inst{1,2} \and Adam J. Finley\inst{3}}

\institute{Rosseland Centre for Solar physics, Universitetet i Oslo, Sem Sælands vei 13, 0371,Oslo, Norway \and Institutt for Teoretisk Astrofysikk, Universitetet i Oslo, Sem Sælands vei 13,0371,Oslo, Norway \and Universit\'e Paris-Saclay, Universit\'e Paris Cit\'e, CEA, CNRS, AIM, 91191, Gif-sur-Yvette, France}

\abstract{Tracing wave activity from the photosphere to the corona has important implications for coronal heating and prediction of the solar wind. Despite extensive theory and simulations, the detection of waves in realistic MHD simulations still presents a large challenge due to wave interaction, mode conversion, and damping mechanisms.}{We conducted this study to detect localised wave activity within a realistic MHD simulation of the solar atmosphere by the Bifrost code. }{We present a new method of detecting the most significant contributions of wave activity within localised areas of the domain, aided by Discrete Fourier Transforms and frequency filtering. We correlate oscillations in the vertical \& horizontal magnetic field, velocities parallel \& perpendicular to the magnetic field, and pressure to infer the nature of the dominant wave modes.}{Our method captures the most powerful frequencies and wavenumbers, as well as providing a new diagnostic for damping processes. We infer the presence of magnetoacoustic waves in the boundaries of prominent chromospheric/coronal swirling features. We find these waves are likely damped by viscous heating in the swirl boundaries, contributing to heating in the upper atmosphere.}{Using the most significant frequencies decomposition, we highlight that energy can be transported from the lower atmosphere to the upper atmosphere through waves and fluctuations along the swirl boundaries. Although further analysis is needed to confirm these findings, our new method provides a path forward to investigate wave activity in the solar atmosphere.}


\maketitle


\section{Introduction} \label{sec:intro}
It has long been debated if the propagation, damping and dissipation of magnetohydrodynamic (MHD) waves is a significant contributor to the coronal heating problem, and if so, where these waves originate from. MHD theory 
\citep{Ferraro_1958,Bel_1971,Nakagawa_1973,Zhugzhda_1983,Zhugzhda_1984,Goedbloed_2004,Goossens_2011} has given us an extensive toolkit to study magnetoacoustic-gravity wave modes, including important insight into how linear and non-linear modes propagate, change, and interact with each other in a non-uniform medium. Specifically, it is important to note the evanescence (damping) of different waves due to criteria known as cut-off frequencies. For example, for acoustic modes this limit depends on the speed of sound, and greatly reduces the purely acoustic wave activity in the upper atmosphere. Furthermore, around the $\beta = 1$ transition, both magnetic and gas pressure can act as the dominant restoring force for magnetoacoustic waves. The distinction of modes in this region is less clear, and it is possible that the nature of the wave may change around this transition, and so mode conversion is usually assumed in this region. In the upper atmosphere, the magnetic field inclination also plays an important role in the trapping and reflection of waves \citep{Nakagawa_1973, Newington_2010}. As the magnetic field becomes more vertical with height, strict constraints develop on the propagation of magnetoacoustic waves into the corona, and most wave energy is instead reflected back into the chromosphere. \\

Whether or not magnetosonic waves have enough energy to heat the corona, it is still unclear how or if they can transport it from the photosphere upwards to the corona. Idealised numerical simulations \citep{Bogdan_2003,Chatterjee_2020,Kumar_kumar_2020,Riedl_acoustic_2021,Riedl_finding_2021,Cally_2022,Yelles_2023} have strengthened our understanding of these modes of oscillation, by tracking and studying driven waves through  inhomogeneous media. Previous studies have shown how energy may be transferred across boundaries via mode conversion, and estimates of wave flux energy have been used to support or disband the significance of specific wave modes in heating the corona \citep{Carlsson_2006,Fossum_2005,Newington_2010,Cally_2017,Tarr_2017, Liu_2023, Mcmurdo_2023}. From this, it is concluded that mainly acoustic modes dominate in the photosphere, whilst magnetoacoustic and Alfvén modes are present in the upper chromosphere and corona. These idealised simulations, however, are still far from the physical reality, lacking crucial physics and self-consistency that may influence wave behaviour. Ultimately, the next steps are to confirm such mode conversions and damping in realistic simulations, and eventually, observations. Furthermore, the estimations of wave energy deposition have important implications for the boundary conditions of solar wind predictions \citep{Holst_2014,Sishtla_2022,Perri2022}. It is possible to estimate the Alfvén wave energy transferred into the solar wind in MHD simulations using the Elssässer variables, and more recently, modified Q-variables that also encompass fast and slow magnetoacoustic wave energy \citep{Van_doorsselaere_2024}. Alternatively, these wave fluxes can be observationally constrained \citep{morton_2012, sharma_2023}, however an accurate determination is a subject of ongoing work. \\

In realistic simulations, wave modes interact with each other, form shocks, experience turbulence, and change nature, making it challenging to follow wave activity through the computational domain. The altitude of the $\beta = 1$ transition is also inhomogeneous, and contains both chromospheric and coronal plasma, such that waves travelling in any direction may cross this transition more than once. Furthermore, when the wave modes are self-driven by convective motions, it is difficult to differentiate individual waves from other time-dependent dynamics. \citet{Yadav_2022} and \citet{Enerhaug_2024} suggest possible methods to detect wave activity across field lines and flux surfaces respectively, by capturing characteristics specific to each magnetoacoustic wave mode. However, these typically do not extract the actual oscillations of wave activity. \citet{Raboonik_2024} proposes a method of decomposition using the eigenenergies of magnetoacoustic waves. This is able to capture both non-linear effects and phase-mixing in 3D realistic simulations, but still has limitations, such as the assumption of adiabaticity (no dissipation mechanisms), and the exclusion of gravity in the derivation of the method. Here, we elaborate on the work of \citet{Finley_2022}, using a combination of Discrete Fourier Transforms (DFT) and component decomposition to infer wave activity in a physically realistic 3D simulation.

\section{Numerical simulation} \label{sec: Implem}
\subsection{The Bifrost code}
The simulation we have analysed was produced using the 3D radiative magneto-hydrodynamic (rMHD) code, \verb|Bifrost|, described in detail in \citet{Gudiksen_2011}. \verb|Bifrost| incorporates a shallow, but self-consistent, convection zone which drives the motions above, up to the mid-corona. Within each atmospheric level, additional physics modules aim to capture the physical processes and create a realistic simulation. Thermal conductivity is applied in the upper atmosphere, and follows Spitzer conductivity, calculated through an explicit hyperbolic approximation \citep{Spitzer1962, Rempel_2017, Cherry_2024}. In optically thin regions (upper chromosphere/corona), radiative transfer is assumed to only depend on density and temperature, with a transfer function calculated with ionization, recombination and collisional excitation rates from precomputed data given by \citet{Shull_1982}, \citet{Arnaud_1985} and \citet{Judge_1994}. Full radiative transfer in four opacity bins \citep{Hayek_2010} is then used for optically thick regions, predominantly in the photosphere and lower chromosphere. These calculations assume a static medium, and local thermal equilibrium (LTE), except for hydrogen ionization in the chromosphere, which is treated as non-LTE. The radiative transfer equation is solved iteratively in order to include important scattering effects. The main MHD equations are solved using a 3rd-order Hyman predictor-corrector scheme, from a staggered Cartesian grid with 5th and 6th order spatial operatives. The simulation was run with variable time-stepping, and localised hyper-diffusive terms to increase stability.
 \begin{figure}
    \centering
    \includegraphics[scale=0.8]{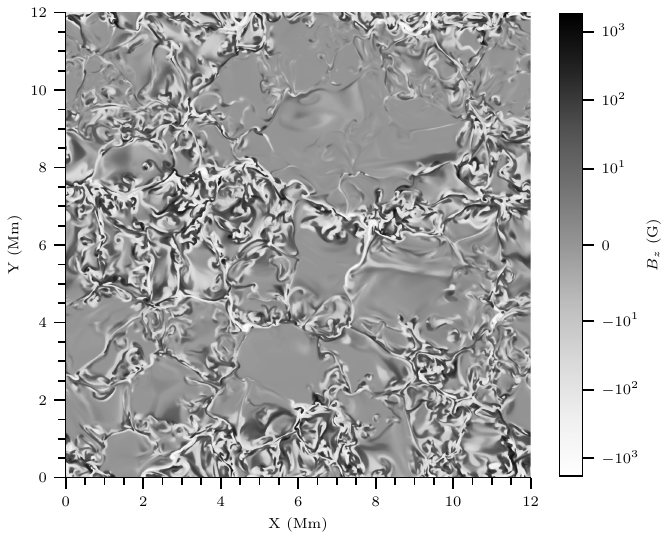}
    \caption{Vertical magnetic configuration where $\tau$ = 1, at time = 136.8 minutes}
    \label{fig: magnetogram}
\end{figure}

\subsection{Set-up}
We analyse a quiet-sun simulation which consists of a $12 \times 12 \, \text{Mm}^2$ horizontal extent, with resolution of $\mathrm{d}x=\mathrm{d}y=11.7$~km. The computational domains extends from the convection zone approximately $2.5$Mm below the photosphere, to the mid-corona, $\sim 8$~Mm above with a varying vertical resolution of $\mathrm{d}z = 6$--$35$~km. In total, the domain consists of $1024^3$ grid points; a higher resolution than most published simulations from \verb|Bifrost|. There are periodic boundaries in $x$ and $y$, with open upper and lower boundaries. The upper boundary uses characteristic boundary conditions (see \citealp{Gudiksen_2011}) which aim to transfer material through the boundary with minimal reflection. Even so, we avoid analysis within 2~Mm of this boundary in order to remove boundary effects. On the lower boundary, the entropy is maintained such that the solar effective temperature reaches 5773 K, although in reality it varies between 5717~K and 5761~K. 
\begin{figure}[h!]
    \centering
    \includegraphics[scale=0.8]{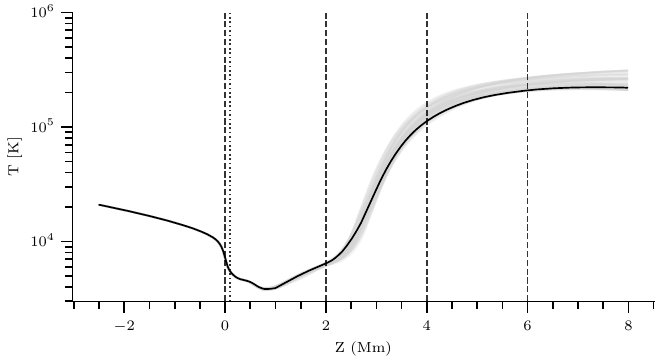}
    \caption{Average vertical temperature profile (black) over the whole domain. The dashed lines display the horizontal planes that were used and analysed in this study, at $z=$ 0, 2, 4, and 6~Mm. The dotted line represents the plane used in Figure \ref{fig: magnetogram}. The grey lines show the evolution of the temperature profile across the analysed timespan. }
    \label{fig: temp}
\end{figure}

\begin{figure*}[h!]
    \centering
    \includegraphics[width=\linewidth]{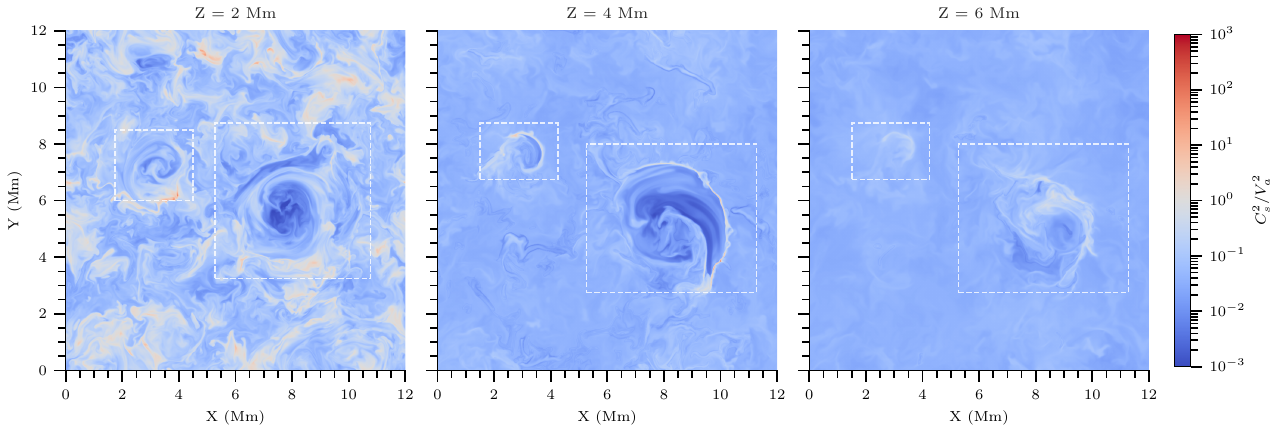}
    \caption{Localised plasma-$\beta$ at height $z= $ 2, 4, and 6~Mm at $t=$~1000 seconds. Regions around $\beta=1$, represented by the lightest shades, are most susceptible to contain phase mixing and damping processes. Two prominent swirling features are highlighted by the boxed regions.}
    \label{fig: beta}
\end{figure*}

This study analyses the simulation starting at 1.5 hours ( referred to as $t=$ 0 seconds) of solar time, with a snapshots available at a cadence of 10 seconds, continuing until 2.1 hours ($t=$ 1890 seconds) of solar time. This pertains to 189 snapshots of data, spanning 36 minutes.

We have chosen three horizontal slices in which to proceed with analysis in this paper, shown in Figure \ref{fig: temp}. These correspond roughly to the upper chromosphere (2~Mm), end of transition region into corona (4~Mm), and the low-corona (6~Mm).

\subsection{Overview}
The simulation evolves a highly non-idealised structure, visible in the vertical magnetic field structure in Figure \ref{fig: magnetogram}. The self-consistent convection zone continually drives the motions further up in the simulation box such that the magnetic field is constantly evolving. The average structure of the atmosphere consists of photospheric granulation at approximately 0~Mm, temperature minimum at 500~km, and the plasma $\beta$,
\begin{equation}
 \beta = \frac{P}{B^2/2\mu_0} \propto \frac{c_s^2}{v_A^2}
 \end{equation}
passes unity ($\beta=1$) in the upper photosphere/lower chromosphere on average (see Figure \ref{fig: beta} for $\beta$ variation across the specific slices). In time and space, the $\beta=1$ level varies considerably between the photosphere, $z=0$~Mm, to the corona, $z>6$~Mm, and forms a highly corrugated surface. Thus, it is not so easy to reach general conclusions about the nature of a wave travelling along or through a horizontal plane. This is especially true for two prominent magnetic features in the domain (see Figure \ref{fig: beta}),  which exhibit torsional "swirling" velocities and twisted magnetic fields, expanding from the upper photosphere into the low corona. The motions of such features are described in depth in \citet{tziotziou_2023}. These structures, henceforth referred to as swirls, elevate the $\beta=1$ layer up to 6 Mm, on the boundaries of the swirl. These thin boundary areas could be important as a possible route for waves to be transferred to the low-$\beta$ plasma from the lower atmosphere. It is also important to note that adiabaticity does not hold everywhere. This is seen by the differences in $p$ and $\rho$ throughout the upper atmosphere, showing that the ideal gas assumption generally doesn't hold. When subject to non-adiabatic conditions, both linear and non-linear waves can lose energy through processes such as damping \citep{Carbonnell2004}.


\section{Methodology}
\label{sec: Methedology}

\begin{figure*}
    \centering
    \includegraphics[scale=0.9]{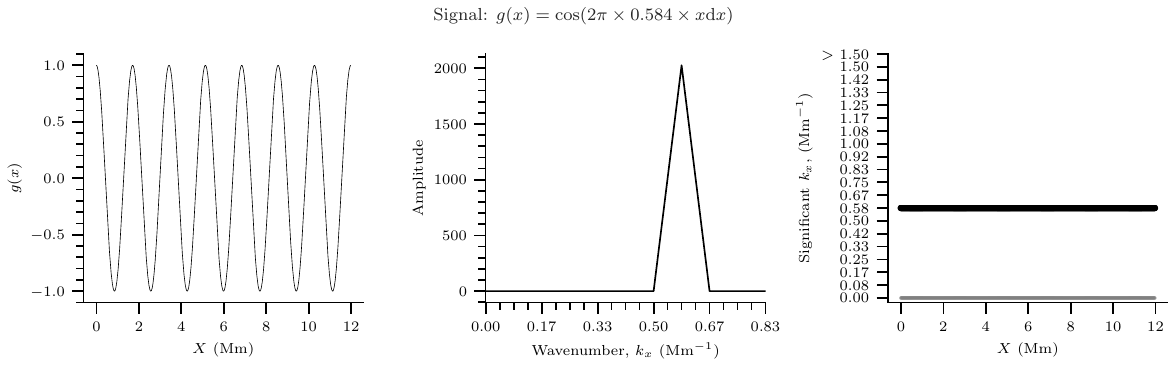}
    \caption{"Proof-of-concept" with an isolated signal in the $xy$-plane, with wavenumber $k_x =$ 0.58~Mm$^{-1}$, and no frequency. The 1st panel shows a 1D slice in the y-axis, taken at an arbitrary time. The middle panel shows the amplitude spectrum from the DFT, averaged across the y and time axes, where there is a sharp peak at the 0.58~Mm$^{-1}$, whilst the third panel shows the most significant signal per pixel in the domain. In grey, the most significant frequency, $\omega$ is shown, and is zero as expected.}
    \label{fig: 1_sig}
\end{figure*}

\begin{figure*}
    \centering
    \includegraphics[scale=0.9]{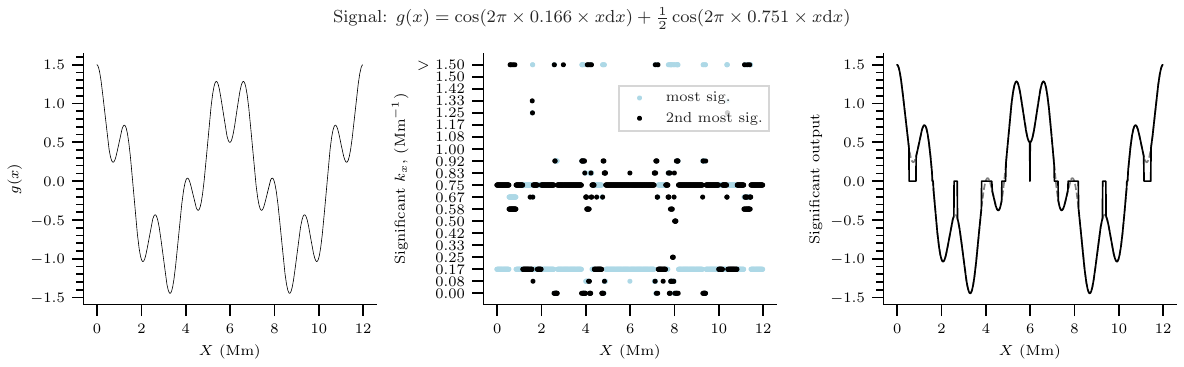}
    \caption{A composite signal of two frequencies, $k_x =$ 0.166~Mm$^{-1}$ and $k_x =$ 0.751~Mm$^{-1}$. The first panel shows the original signal, the second panel shows the results of the most and second most significant calculations, and the final panel shows the final output of the significant frequencies. The original frequency is shown by the dotted lines for reference. }
    \label{fig: 2_sigs}
\end{figure*}
Inspired by the analysis in \citet{Finley_2022}, we use the DFT on the temporal and spatial evolution of a quantity, $g(x,y,t)$, in a horizontal plane. The quantity is mapped to Fourier space, in terms of the horizontal wavenumbers, $\tilde{\nu_i} = k_i/2\pi$, and the frequencies, $f = \omega/2\pi$,
\begin{multline}
     \mathcal{F}(\tilde{\nu}_x,\tilde{\nu}_y,f) = \\
    \sum_{n_x = 1}^{N_x} \left(\sum_{n_y = 1}^{N_y} \left(\sum_{n_t = 1}^{N_t} g(n_x,n_y,n_t)e^{-i \tfrac{ 2\pi f n_t}{N_t}} \right) e^{-i \tfrac{ 2\pi \tilde{\nu}_x n_x}{N_x}} \right) e^{-i\tfrac{2\pi \tilde{\nu}_y n_y}{N_y}},    
\end{multline}
where $n_x$,$n_y$,$n_t$, represent the indices of the spatial and temporal directions, and $N_x,N_y,N_t$ denote the total number of points in each direction. The range of the spatial and temporal frequencies are given by the spatial and temporal step-size, respectively, for the maximum frequency, and the size of the spatial/temporal domain for the minimum frequency. Therefore, the minimum frequency sampled is 1/L = 0.083 Mm$^{-1}$ for the spatial directions, and 1/T = 0.52 mHz for the temporal direction. These are also the respective sampling resolutions. \\

For a given set of filtered horizontal wavenumbers or temporal frequencies (frequency bin), the localised contributions to the quantity, $g$, in the original spatial-temporal domain may be reconstructed  using an inverse DFT. We propose that localised wave activity may be detected in regions of the simulation domain by mapping the frequencies that reconstruct the most accurate strength at each point. Per index, the significance of a frequency bin, $b$, is determined through the absolute difference of the reconstructed signal, $S_b$, to the original signal,
\begin{equation}
    \text{Diff}_b[n_x,n_y,n_t] = | \,S_b[n_x,n_y,n_t] - g[n_x,n_y,n_t]\,|. \label{eqn:rel_err}
\end{equation}
Therefore, the most significant signal, has the smallest absolute error. We note that if the signal is not centred on zero, it is important to combine the zeroth frequency bin with the other individual bins, in order to accurately compare the signals.\\

Figure \ref{fig: 1_sig} displays the some-what trivial result for a signal with a single frequency, where the most significant frequency (MSF) calculation is comparing individual frequencies. When there is only one dominating frequency, the MSF will be exactly the input frequency, since all other frequencies will have zero amplitudes. When two or more waves travel through a domain, it is important to repeat the MSF calculation to capture the second-most significant frequency (SSF), as seen in Figure \ref{fig: 2_sigs}. Here, the signal is comprised of two frequencies, the main wave has a large amplitude and low frequency whilst a smaller-amplitude higher-frequency wave runs across it, creating a more complex signal. The MSF calculation predominantly captures the low frequency wave. To repeat this process for the SSFs, we remove the signal produced by the composite MSFs from the original signal, leaving the point-wise residual, $\text{Diff}_{MSF}$. The MSF algorithm may then be applied to the residual, resulting in the SSF. We can see from the 3rd panel of Figure \ref{fig: 2_sigs}, that the output of the MSF and SSF signals combined returns a signal very similar to the original, except for in a few places around turning points. At these turning points, the distribution of MSFs is found to be either the zero frequency (relating to the equilibrium value), or the frequency bin that contains all frequencies higher than 1.50~Mm$^{-1}$, which could relate to numerical noise in the signal. We note that the spread of frequencies detected in each MSF calculation may vary dependent on the difference of frequency and amplitude between the two waves, but in any case, after two calculations, we find that the two dominant frequencies are clear. \\

This method allows us to go beyond the standard DFT power spectrum in two ways. Firstly, it can show where a frequency becomes dominant in a specific region of the simulation domain where the dynamics are continuously changing. Secondly, it allows us to create inverse $\omega$-frequency outputs in a spatial plane, and vice-versa, allowing us to locate regions of the domain space that contain fluctuations at certain frequencies and wavenumbers.

\subsection{Damping}
\begin{figure}[h!]
    \sidecaption
    \includegraphics[width=0.9\linewidth]{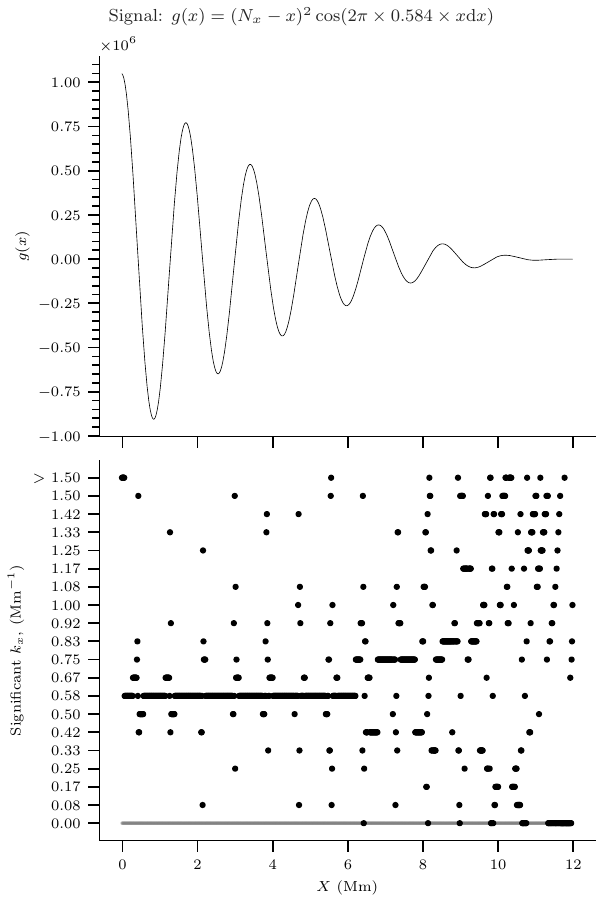}
    \caption{Most significant frequencies calculated for a signal with a damped amplitude.}
    \label{fig: Damped_sig}
\end{figure}

It is a general characteristic of DFTs that they only decompose a signal into frequencies with constant amplitudes. Therefore, it is difficult to capture decaying signals, such as the ones imperative to energy transfer in the upper solar atmosphere, through damping and dissipation. In Figure \ref{fig: Damped_sig}, we show how damped signals can be detected by calculating the MSFs. When the amplitude of the wave is large, the MSF is confidently set to the correct frequency, but as the amplitude diminishes, the spread of "significant" frequencies increases, with the power shifting away from the actual frequency, and towards the lowest (0~Mm$^{-1}$) and highest ($> 1.5 \text{ Mm}^{-1}$) frequencies. This is a distinct characteristic for a damped wave: therefore we can infer the damping of signals in regions where a strong signals dissolves into pixelated high and low frequencies.

\section{Results}
We present a series of results that demonstrate the effectiveness of this method for capturing localised wave activity in realistic simulations. We note that future work will explore the physical mechanisms at play for each case (Cherry, Gudiksen, \& Finley, in prep.), since they require more in-depth analysis that is outside the scope of this paper.

\subsection{Power Spectral Density and pixel plots} \label{sec: general_trends}

\begin{figure}
    \centering
    \includegraphics[width=\linewidth]{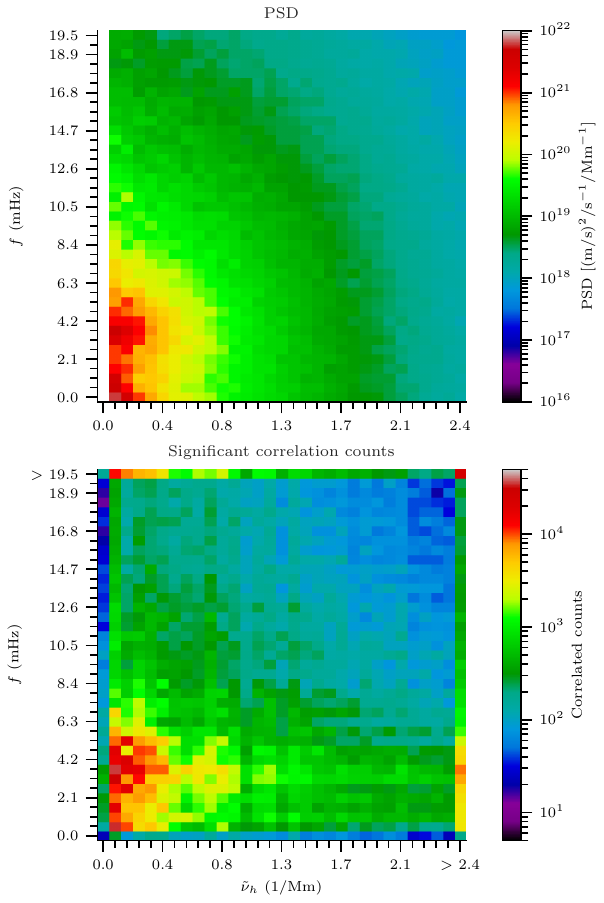}
    \caption{Power Spectral Density plot for $u_\parallel$ (right), and corresponding MSF counts (left). Taken for the horizontal plane $z =$ 4~Mm. On the right plot, the outermost values represent the total counts for all other frequencies.}
    \label{fig: PSD}
\end{figure}

Usually, the results of a DFT are displayed as a Power Spectral Density (PSD), where
\begin{equation}
    \text{PSD}( \tilde{\nu}_x,\tilde{\nu}_y,f) = |\mathcal{F}(\tilde{\nu}_x, \tilde{\nu}_y,f)|^2
\end{equation}
shows which frequencies contain the most power over the entire domain. High values of the PSD either correspond to the sum of many fluctuations covering a large area of the domain, or more localised fluctuations with large amplitudes in the domain. Figure \ref{fig: PSD} concerns the velocity parallel to the magnetic field,
\begin{equation}
    u_\parallel = \mathbf{u} \cdot \mathbf{B},
\end{equation}
at $z=4$~Mm, and shows the frequency against the horizontal wavenumber,
\begin{equation}
\tilde{\nu}_h = (\tilde{\nu}_x^2 + \tilde{\nu}_y^2)^{1/2}.    
\end{equation}
The DFT PSD (top panel) is compared to an equivalent plot for the MSF method (bottom panel). For this, we count the number of pixels with each combination of MSF and most significant wavenumber (MSW) for a specific snapshot. In all examples in this paper unless explicitly stated, we have taken a snapshot at time $t=1000$~seconds of the analysed data and considered each discrete frequency sampled by the DFT individually (each frequency bin contains only one frequency, with width 0.525~mHz). In the case of the perpendicular wavenumber, the bins are created such that each wavenumber bin is centred on the individual wavenumbers in $x$, $\tilde{\nu}_x$, sampled by the DFT. Each bin therefore contains 2-3 wavenumbers, and has a width of 0.08~Mm$^{-1}$. \\

\begin{figure*}[h!]
    \centering
    \includegraphics[width=\linewidth]{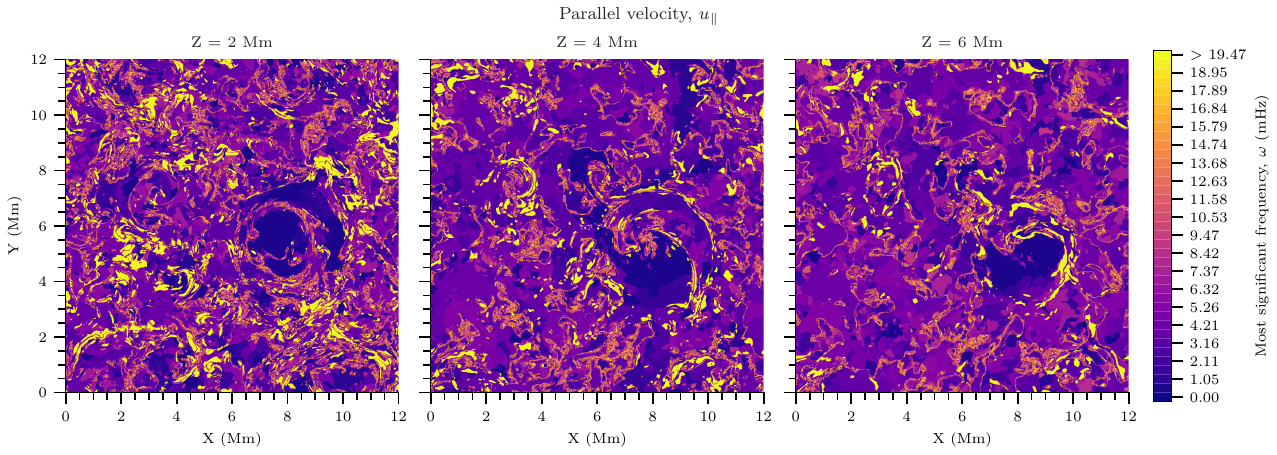}
    \caption{MSF in the velocity parallel to the magnetic field, $u_\parallel$, for the chromosphere and upper atmosphere.}
    \label{fig: upar}
\end{figure*}

\begin{figure*}[h!]
    \centering
    \includegraphics[width=\linewidth]{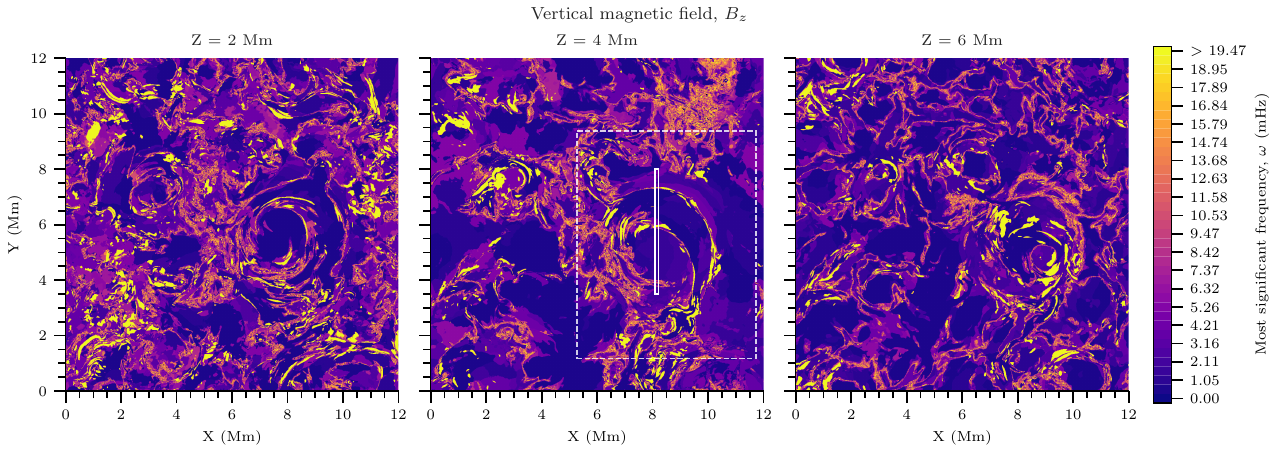}
    \caption{MSF for the vertical magnetic field, $B_z$, in the chromosphere and upper atmosphere. The rectangles represent the locations shown in later figures:  the dashed rectangle relates to Figure \ref{fig: Damped_1D}, and the solid line represents the slice in Figure \ref{fig: swirl}.}
    \label{fig: B}
\end{figure*}

The MSF method captures the most powerful areas of the PSD in Figure \ref{fig: PSD}, confirming that the method works on the most important fluctuations in the domain. In the mid-power range, the MSF method highlights a group of larger wave numbers with frequency around 4~mHz whose importance is obscured by the PSD. The right-most column ($\tilde{\nu}_h > 2.4\text{ Mm}^{-1}$), and similarly the upper-most row ($f > 19.5 \text{ mHz}$) also show strong signals. However, this is because they contain all the higher frequencies (up to 50~mHz) and wavenumbers (up to 42.6~Mm$^{-1}$), respectively, which are not considered individually, and instead the associated power stacks up in a single bin. The advantage of the MSF calculations is that a pixel plot, such as in Figure \ref{fig: upar}, may be used to show the location of the MSFs in the domain, and by extension, the nature by which the power of the PSD is created. Figure \ref{fig: upar} highlights the areas where fluctuations of a specific frequency begin to dominate in the domain, either locally or globally. For example, we see an overwhelming development of MSFs in the range 3-6~mHz as height increases. By $z=4$~Mm, there are expansive and continuous patches where these frequencies dominate, covering $45\%$ of the domain. This compliments the results from the PSD in Figure \ref{fig: PSD} that the most power is centred on these frequencies. By $z=6$~Mm, the coverage has reached $50\%$, whilst in the photosphere (not shown) it is only $\sim 25 \%$. The same activity is not observed in the magnetic field in Figure \ref{fig: B}, and so it is likely these fluctuations relate to longitudinal pressure modes. Although outside the scope of this paper, we note that since we now have location information, it would be possible to estimate the wave flux energy more accurately, by only integrating over the regions covered by the specific frequencies, instead of just taking an average over the whole domain. Thus the wave-energy for longitudinal waves would be calculated by,
\begin{equation}
    F_\text{long} = \frac{1}{2}\iint_\mathbf{S} \rho v_p^2 c_s \, \mathrm{d}\mathbf{S},
\end{equation}
where $v_p$ is the velocity in the direction of propagation, and $\mathbf{S}$ is the area covered by the specific frequencies. Since the direction of propagation is unknown, an upper limit could be estimated with the magnitude of the velocity vector, $|\mathbf{v}|$.

\subsection{Matched frequencies} 
\label{sec: matched}
\begin{figure}
    \centering
    \includegraphics[width=\linewidth]{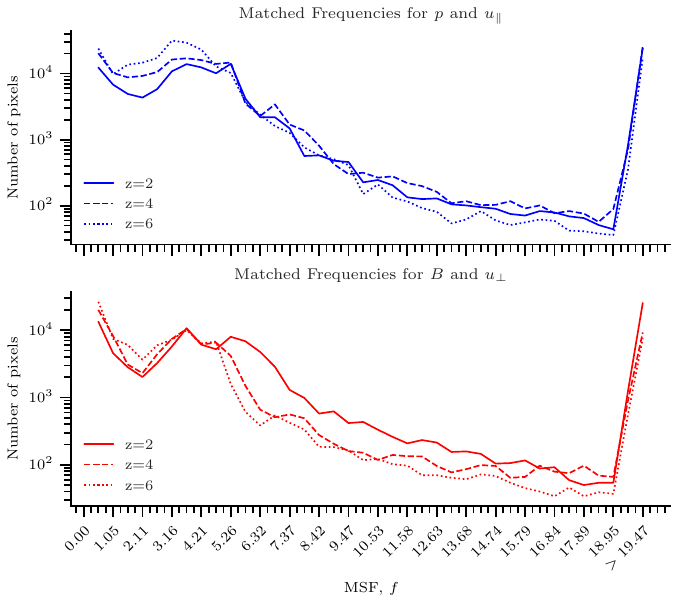}
    \caption{Number of matched MSFs between pressure \& parallel velocity (upper), and magnetic field \& perpendicular velocity (lower), at $z=4$~Mm. The frequencies are considered matched if they are within $\pm \sim 0.53$~mHz (1 frequency step) of each other.}
    \label{fig: corr_counts_diag}
\end{figure}

\begin{figure*}
    \centering
    \includegraphics[width=\linewidth]{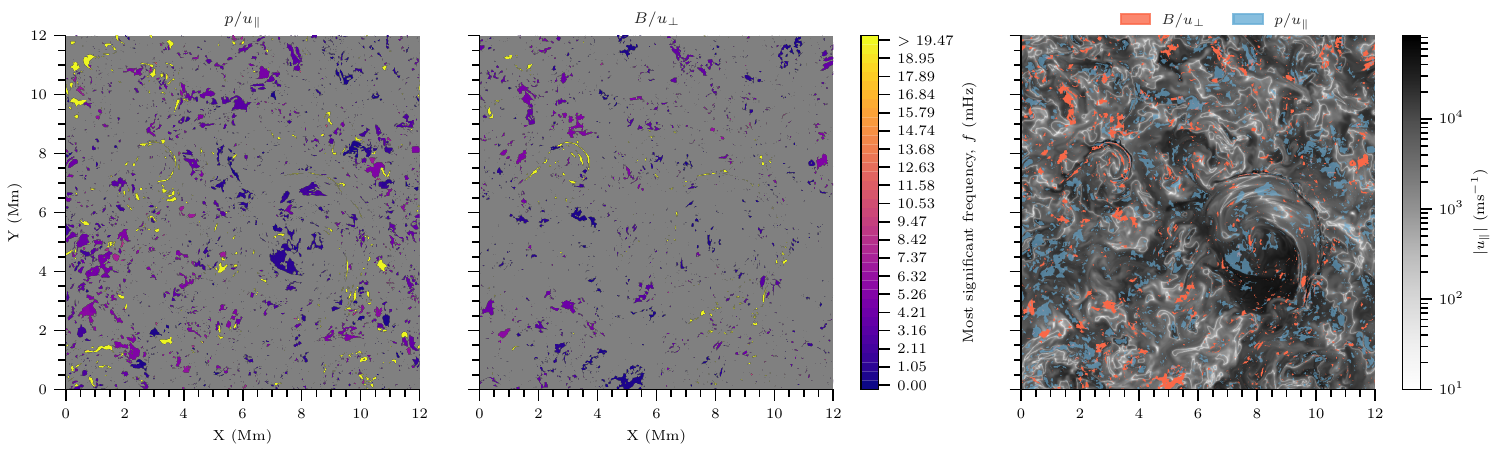}
    \caption{Regions of matched MSFs between the magnetic field \& perpendicular velocity (left) and the gas pressure \& parallel velocity (middle) at $z=4$~Mm. The frequencies are considered matched if they are within $\pm \sim 0.53$~mHz (1 frequency step) of each other.
    The right panel shows the results from the left two panels overlayed to show overlapping regions. "Magnetic" fluctuations are shown in red, whilst "acoustic" fluctuations are in blue. Here, the magnitude of the frequencies is omitted. The magnitude of parallel velocity is shown for reference in the background. For all plots, MSF's of 0~mHz have been removed as a trivial result.}
    \label{fig: matched_frequencies}
\end{figure*}

 We can speculate on the nature of other fluctuations using the correlations between MSFs in the magnetic field, parallel and perpendicular velocity, and pressure. We use the example of magnetoacoustic modes in the upper atmosphere. Magnetoacoustic modes have both pressure and magnetic variations, as well as transverse and longitudinal oscillations. For the longitudinal component, we examine the parallel velocity and pressure together. The top panel of Figure \ref{fig: corr_counts_diag} 
 counts the number of pixels where the MSF is the same for both quantities, $\pm$ one frequency step (0.53~mHz). The strongest correlations between the two variables always lie within the $3-6$~mHz range, supporting the idea that the dominating coverage in Figure \ref{fig: upar} is due to pressure modes. There is also a clear correlation in this frequency range for the transverse oscillations, represented by the magnitude of the magnetic field and perpendicular velocity, $v_\perp$ (lower panels of Figure \ref{fig: corr_counts_diag}). From Figures \ref{fig: upar} and \ref{fig: B}, however, we already know that the locations of fluctuations with these frequencies in the magnetic field and parallel velocity in general do not match in location, and this is confirmed in Figure \ref{fig: matched_frequencies}. It is an interesting feature that we nevertheless see a similar number of pixels showing transverse oscillations as longitudinal in this region if they do not lie in similar locations. \\
 
 Using the correlated pixel plots in Figure \ref{fig: matched_frequencies}, we can distinguish areas where the magnetic and acoustic MSFs do overlap. Matching the results of all four variables is an ambitious task, and we would expect that the less dominant oscillations (of fast- and slow-magnetoacoustic modes) would more likely be captured in the secondary, or even tertiary iterations of the MSF calculations. Considering this, it is not unexpected that there are only a few overlapping regions to be found in Figure \ref{fig: matched_frequencies}. Nevertheless, there is a notable coupling between both longitudinal and transverse components on the boundaries of both swirls, suggesting the presence of magnetoacoustic wave modes in these areas. We note that these regions do not have weak magnitudes in either velocities or pressure relative to the average of the domain, as shown for parallel velocity in the right-most panel of Figure \ref{fig: matched_frequencies}). Therefore, these results are unlikely to be affected by noise or other artefacts from the method. On the other hand, the magnitude of the magnetic field can be low in these areas. Even so, the matching of three strong-valued variables inspires further analysis.
 
\subsection{Small-scale processes} \label{sec: small-scale}

\begin{figure*}[h!]
    \centering
    \includegraphics[scale=0.8]{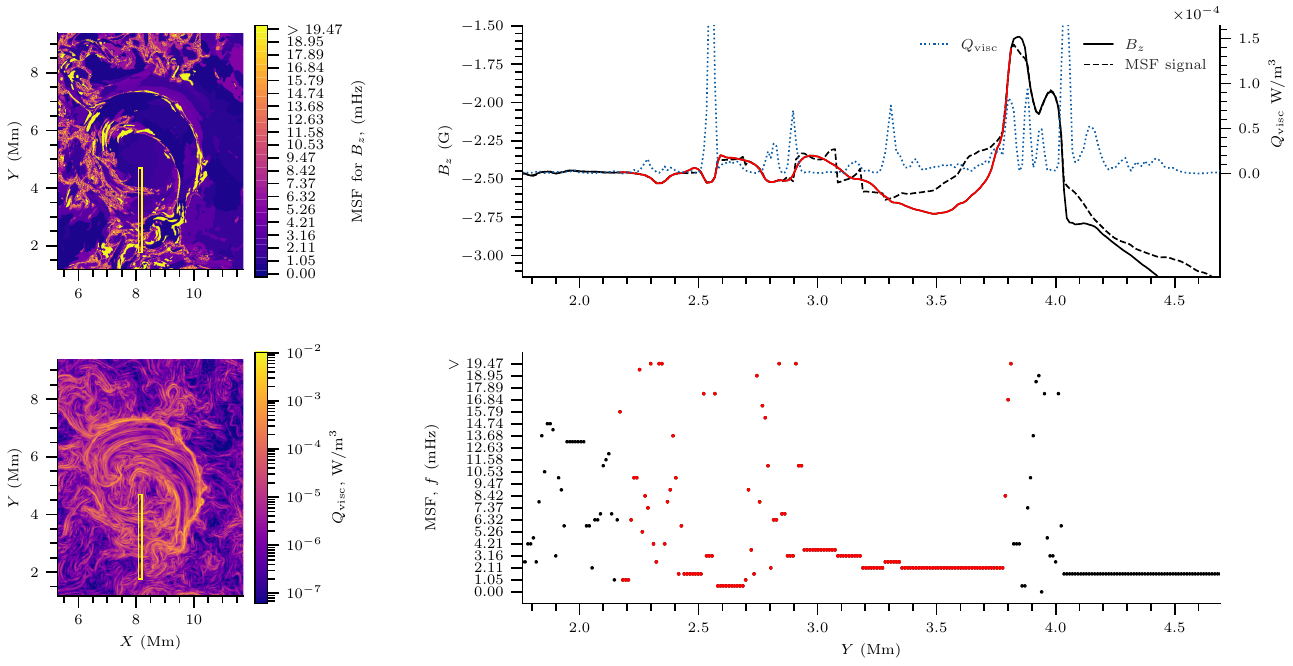}
    \caption{A 1D slice at $X \approx 8.20$~Mm through the vertical magnetic field, $B_z$, at $z=4$~Mm shows that the speckled high-frequencies capture wave dissipation across the swirl boundary between $Y\approx 1.76$~Mm and $Y\approx 4.69$~Mm. The viscous heating, $Q_\text{visc}$, also correlates with this conclusion. The sliced region is emphasised by the yellow box in the left panels, and the damped oscillation region is highlighted in red on the right panels.}
    \label{fig: Damped_1D}
\end{figure*}

In contrast to the MSFs of the parallel velocity, the vertical magnetic field, shown in Figure \ref{fig: B}, shows global fluctuations at the lowest frequencies in the upper atmosphere. We see concentrations of fluctuations with a broad range of frequencies localised along the magnetic field boundaries in the domain. In fact, for the $3-6$~mHz frequencies in $B_z$, the coverage steadily decreases from $\sim 30\%$ at z$z=2$~Mm, to $\sim 20\%$ by $z=6$~Mm, with strong concentrations around the edges of both swirls at $z=4$~Mm. The behaviour of these concentrations are readily identified using the MSF method as it separates structure in the spatial and frequency domains. \\

Furthermore, lines of "speckled" high-frequency MSFs are present in both the parallel velocity and vertical magnetic field, seen in the orange-yellow range in Figure \ref{fig: B}.  These networks of higher-frequencies are observed in many of the variables, but most structured in the vertical magnetic field, $B_z$, along the edges of the swirls. In these regions, the MSF changes over very short distances ($\sim$ per pixel) in the spatial domain, creating the "speckled" pattern. It is unlikely that this is a physical phenomenon, and instead a "short-coming" of the method, which we can exploit to detect other phenomena. There are a number of reasons these patterns could arise. Firstly, in regions where the original signal is close to zero relative to the average of the domain, it is important to acknowledge the presence of noise in the simulation. The noise is most likely to be captured by the top "highest frequencies" bin (i.e., displayed in yellow), although it is also possible in the high individual frequency bins. However, around the edge of the swirls, for example, the magnetic field is only near zero in very small regions where the polarity changes, and the velocities remain strong, so the presence of these speckled regions is likely to be the result of a different phenomenon. Instead,  we refer to the behaviour discussed in Section \ref{sec: Methedology}. Aside from noise, these regions could either indicate a smaller, secondary fluctuation with different frequency, as described in Figure \ref{fig: 2_sigs}, where the MSF method sometimes also captures the SSF, with some loss of precision, or it indicates the dissipation of a wave, as in Figure \ref{fig: Damped_sig}. There is a slight difference to the idealised example of the latter used in Section \ref{sec: Methedology}: in this case we are taking the temporal frequency, $f$, and relating it to spatial location, and so we rely on the assumption that the MSFs will show damping across the orthogonal direction in a similar fashion to how they react in the direction parallel to the given frequency.\\

In the case of the thicker areas of "speckled" MSFs for $B_z$ on the boundary of the large swirl, at $z=4$~Mm, the SSFs (not shown) provide no evidence of a secondary fluctuation in these regions. Further analysis favours the dissipation theory:  the $B_z$ signal appears to be damped in a quasi-oscillatory motion from the centre of the swirl, at $\sim$ (8~Mm,5~Mm), to the outside boundary located vertically downwards at $Y\approx 2$~Mm. This is displayed in the top-right panel of Figure \ref{fig: Damped_1D} as a 1D vertical slice through the lower part of the swirl. The bottom right panel shows that this observed oscillation coincides with chaotic switching of MSFs, in a similar fashion to the distinct pattern for damping described in Section \ref{sec: Methedology}. In this region we also see several peaks in the viscous heating term, $Q_\text{visc}$, which measures the transfer of energy from kinetic to heat in the plasma through viscous forces. There is an increased spatial-average across the slice compared to other areas in the domain, as seen in the lower-left panel. In the region of the observed dissipation, the peaks align exactly with each dip in the $B_z$ oscillation, suggesting energy is transferred from the wave into heat in this region. In contrast, other regions close to the swirl boundary display a thinner region of much stronger viscous heating, coinciding with the thin "speckled" regions. In these regions, we do not see oscillations, but only a sharp decrease in the magnitude of $B_z$ to near zero. This observation, coupled with the strong viscous heating suggests that the damping rate here is much greater than the frequency of the wave, such that no oscillations are observed. \\

The above example demonstrates the kind of further analysis that is needed in order to distinguish the nature of any "speckled" region observed. It is quite likely that there are more instances of damping to be found in this simulation. Similarly, it would be possible to track these damping mechanisms through the atmosphere. If the above speculations are validated through further analysis, this could provide an invaluable tool for measuring the contribution to coronal heating from the dissipation of waves.

\section{Discussion}\label{sec:concl}
\subsection{Viscous heating in swirl boundaries}
Despite this paper primarily serving as an introduction to the MSF method, the results from Sections \ref{sec: matched} and \ref{sec: small-scale} can be used to understand how energy is transported via wave modes and deposited in the low corona. In the swirl boundaries, the plasma-$\beta$ remains greater than 1, as seen in Figure \ref{fig: beta}. Therefore the plasma displays characteristics of chromospheric plasma, but is pulled up to coronal heights by the motions within the swirls. Although greater than 1, the plasma-$\beta$ remains close to unity, providing ideal conditions in the plasma for mode conversion, via phase-mixing or otherwise. This allows waves with both strong magnetic and acoustic characteristics to exist. As the oscillations diffuse outwards from the swirl boundaries, the plasma approaches low-$\beta$ once more, and so the conditions for magnetosonic waves to propagate change, causing damping of the wave amplitude from external forces such as viscosity. The excess energy is transferred into heat, and found in the viscous heating term, $Q_\text{visc}$, that ultimately raises the temperature of the plasma. This conclusion requires further analysis to validate, namely by following plasma along the spiral boundary and "tracking" specific waves. This method should provide the tools to do this in future (Cherry et al. in prep).

\subsection{Contributions of the MSFs}

\begin{figure}
    \centering
    \includegraphics[scale=0.8]{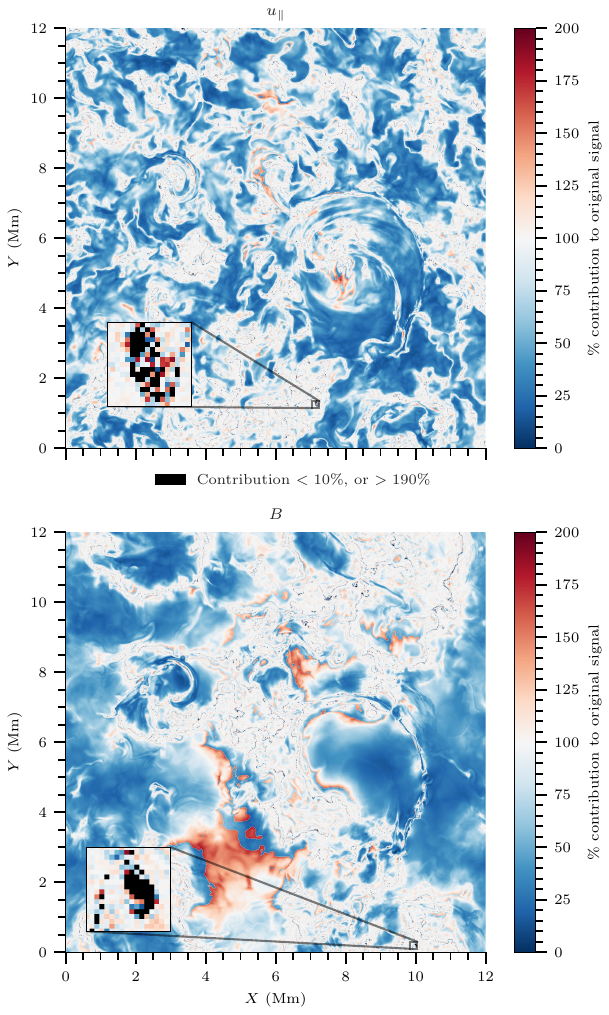}
    \caption{The contribution of the MSF signal to the original signal of parallel velocity (upper) and magnetic field (lower). The insets provide examples of the small patches of negligible contributions by the MSF signal.}
    \label{fig: Error}
\end{figure}

Figure \ref{fig: Error} shows the contribution,
\begin{equation}
    C = (1-\text{Diff}_{\text{MSF}}) \times 100 \%,
\end{equation}
of the MSF signal to the original signal, where $\text{Diff}_\text{MSF}$ is the difference between the two signals, calculated in the same way as in Equation \ref{eqn:rel_err}, but including the sign of the difference. We do not expect one signal to account for the entire strength of the original signal, due to the extent at which neighbouring processes are continuously interacting. In theory, both a contribution greater than 100\% and a negative contribution are possible, since the MSF signal may overestimate the original signal, or be counteracting the effect an overestimated signal through destructive interference (with opposite sign to the original signal). In our results, there are no negative contributions. This is because in most cases, the overestimated signal will be closer to the original than the counteracting signal. The likelihood the method provides an accurate result is lower for areas where the contribution is further from 100 \% (i.e., close to 0\% or 200\%). Nevertheless, Figure \ref{fig: Error} shows there are only a few areas where this is the case (shown by the black patches), and that these areas do not coincide with the locations of the results addressed in this study. \\ 

\begin{figure}
    \centering
    \includegraphics[width=\linewidth]{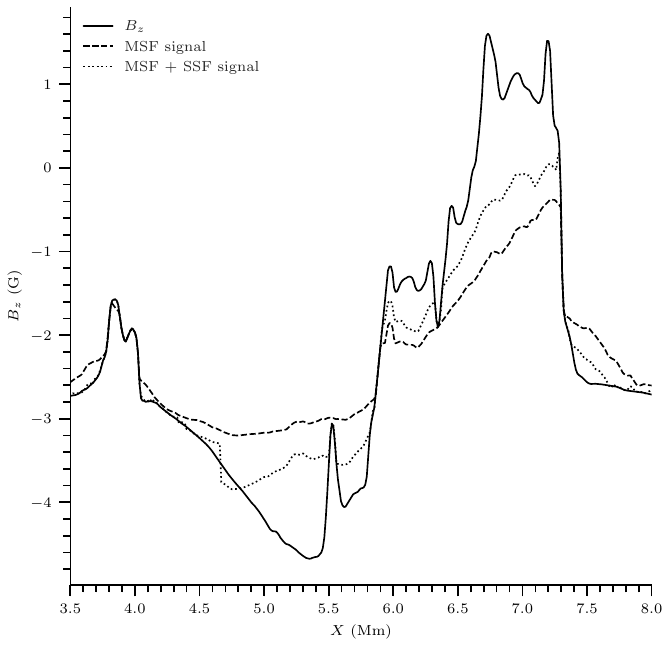}
    \caption{The same 1D slice as in Figure \ref{fig: Damped_1D}, at $X \approx 8.20$~Mm, for the vertical $B_z$ across the largest swirl. The signal created from the MSFs is shown by the dashed line, whilst the signal created when the SSFs are added is represented by the dotted line. The oscillations along the peak from $X \approx 6$~Mm to $X \approx 7.5$~Mm are not captured by either signal.}
    \label{fig: swirl}
\end{figure}

For structures where the original signal at a specific time is much larger than the average deviation, such as in the strong velocities and magnetic field of the swirls (that move in space and time), the MSF method can accurately depict the spatial and time scales, but not necessarily the correct contribution of the MSF or MSW signal to the original. Figure \ref{fig: swirl} is an example of this for the large swirl. This is because the MSF relies on the constant amplitudes taken from the DFT, which will be an average of the fluctuations in the domain for any given wavenumber. Therefore it is not the purpose of the MSF method to detect accurately the contribution of large, time-evolving structures in a global domain. If this is required, one could apply the DFT to a smaller region around the structure in question.

\subsection{Future applications}

In this paper, we have established an algorithm for detecting localised wave activity based on the results of DFTs. This method provides a new way to probe wave activity in the simulation domain, and has already produced results that show potential for improving our understanding of heating in the solar atmosphere. It surpasses the global results given by a DFTs power spectrum, by providing information on dominant frequencies in localised areas. This gives us valuable insight into the small-scale activity surrounding areas of interest such as swirl boundaries and the $\beta=1$ transition. Furthermore, the behaviour of the MSFs for damped oscillations may be exploited for detection of damping mechanisms, which is not possible in the DFT alone. The algorithm is an alternative to the Wavelet method \citep{jess2023} which does not require as many assumptions about the wave modes. It gives more precise results of frequency and location than decomposition methods such as Empirical Mode Decomposition \citep{Huang1998} and Single value decomposition \citep{Santolik2003}.  \\

We have focused our attention on the MSFs and MSWs in the horizontal spatial domain.
In the upper atmosphere, the majority of oscillations propagate vertically upwards, driven by the convective motions below, and more analysis in the vertical direction is an important addition to future work. This could help to refine the results found on the boundaries of the swirl. A 3D MSF analysis could be used to find the direction of propagation in the boundary regions, and track fluctuation origin, propagation, and dissipation in time and as a function of height. The results from which could be compared to those of traditional wave guides, such as described by \citet{Enerhaug_2024}. A similar application would be detailing the wave activity along coronal loops in a 3D simulation. This could extend the work of, for example, \citep{Riedl_acoustic_2021} by including external physics in the plasma around the loop. Advanced 3+-dimensional calculations require significant memory and computational power due to the highly-resolved spatio-temporal coverage in this simulation. By adapting the DFTs or domain-space used, we could reduce this requirement, which could also improve the method in areas of large structures, as discussed above. Further analysis of the MSFs and their distribution in the domain will give us valuable insights into the origin and heating of the solar wind.

\begin{acknowledgements}
This project has received funding from the European Union's Horizon 2020 research and innovation programme under the Marie Skłodowska-Curie [grant agreement Nº 945371]. It is also supported by the Research Council of Norway through its Centres of Excellence scheme, project number 262622, and through grants of computing time from the Programme for Supercomputing.

A.J.F received funding from the European Research Council (ERC) under the European Union’s Horizon 2020 research and innovation programme (grant agreement No 810218 WHOLESUN).

\end{acknowledgements}

\bibliographystyle{aa}
\bibliography{Main}

\end{document}